\documentstyle[psfig,aps,twocolumn]{revtex}
\def\be{\begin{eqnarray}}
\def\ee{\end{eqnarray}}

\def\la{\langle}
\def\ra{\rangle}

\def\roughly#1{\mathrel{\raise.3ex\hbox{$#1$\kern-.75em%
\lower1ex\hbox{$\sim$}}}}

\newcommand{\eq}{\begin{equation}}
\newcommand{\eqx}{\end{equation}}
\newcommand{\eqn}{\begin{eqnarray}}
\newcommand{\eqnx}{\end{eqnarray}}
\newcommand{\f}[2]{\frac{#1}{#2}}
\newcommand{\lm}{\lambda}
\newcommand{\Lm}{\Lambda}

\newcommand{\al}{\alpha}

\newcommand{\eps}{\varepsilon}
\newcommand{\dl}{\delta}

\begin{document}

\renewcommand{\thefootnote}{\arabic{footnote}}
\setcounter{footnote}{0}

\vskip 0.4cm

\title{\LARGE\bf Free Random L\'{e}vy Matrices}

\author{
Zdzis\l{}aw Burda$^{a,b}$\thanks{E-mail: burda@physik.uni-bielefeld.de}, 
Romuald A. Janik$^{a}$\thanks{E-mail: ufrjanik@jetta.if.uj.edu.pl},
Jerzy Jurkiewicz$^{a}$\thanks{E-mail: jurkiewi@hetws14.nbi.dk},
Maciej A. Nowak$^{a,c,d}$\thanks{E-mail: nowak@kiwi.if.uj.edu.pl },\\
Gabor Papp$^{c,e}$\thanks{E-mail: pg@ludens.elte.hu}
and Ismail Zahed$^{c}$\thanks{E-mail: zahed@zahed.physics.sunysb.edu}}

\address{
$^a$
{\it M. Smoluchowski Institute of Physics,
Jagellonian University, Cracow, Poland} \\
$^b$
{\it Fakult\"at f\"ur Physik, Universit\"at Bielefeld
P.O.Box 100131, D-33501 Bielefeld, Germany} \\
$^c$
{\it Department of Physics and Astronomy,
SUNY-Stony-Brook, NY 11794  U.\,S.\,A.} \\
$^d$
{\it Physics Department, Brookhaven National Laboratory,
Upton, NY 11973, U.\,S.\,A.} \\
$^e$
{\it HAS Research Group for Theoretical Physics, 
E\"otv\"os University, Budapest, H-1518 Hungary}
}

\date{\today}
\maketitle
\begin{abstract}
Using the theory of free random variables (FRV) and the Coulomb gas
analogy, we construct stable random matrix ensembles that are random 
matrix generalizations of the classical one-dimensional stable L\'{e}vy 
distributions. We show that the  resolvents for the corresponding
matrices obey transcendental equations in the large size limit. 
We solve these equations in a number of 
cases, and show that the eigenvalue distributions exhibit L\'{e}vy tails. 
For the analytically known L\'{e}vy measures we explicitly construct the
density of states using the method of orthogonal polynomials. We show
that  the L\'{e}vy tail-distributions are characterized by a novel form 
of microscopic universality.
\end{abstract}



\section{Introduction}
There is a wide  and 
 growing interest in stochastic processes with
  long tails in relation to self-similar phenomena.
Long tails cause the 2D random walk to be attracted to
L\'{e}vy stable fixed points with ubiquitous physical properties such as
intermittent behavior and anomalous diffusion. Physical examples
include charge carrier transport in amorphous semiconductors, vortex
motion in high temperature 
superconductors, moving interfaces in  porous media, 
spin glasses and anomalous heat flow in heavy ion
collisions. L\'{e}vy distributions have also find
applications in biophysics, health-physics and finances.
Their importance stems from their stability
under convolution, i.e. the sum of two L\'{e}vy distributed random
variables follows also a  L\'{e}vy distribution. L\'{e}vy stability
for scale-free processes is the analogue of Gauss-stability for 
scale-dependent processes with finite variance.

In the realm of complex and/or disordered systems, 
the theory of random  matrices plays an important role in differentiating
noise from information. Also, it allows for a generic analysis of
complex phenomena in the chaotic regime using random matrix
universality. So far, there does not seem to be a 
developed theory of random L\'{e}vy matrices, with the exception
of~\cite{BC} which is not based on FRV. The reasons are two-fold: L\'{e}vy
distributions are usually defined by their characteristic functions, 
making their probability density functions (pdf) usually user-unfriendly.
Second, L\'{e}vy distributions do not have finite second and higher
moments, making standard  techniques of random matrix theory usually of 
little use.

The aim of this paper is to provide an appropriate generalization of
the classical stable one-dimensional L\'{e}vy distributions to the
random matrix setting. We use results from the theory of free random variables,
to formulate a theory of stable L\'{e}vy random matrices. 
In section~2, we use the explicit results from 
Bercovici and Voiculescu~\cite{BV} to show that  the 
resolvent for free random variables obeys certain transcendental equations. 
In section~3, we argue that the formal result of FRV 
could be explicitly constructed using large matrices. 
We use the Coulomb gas analogy to construct  the L\'{e}vy measures.
In section~4, we  show how the orthogonal polynomial method can be
applied to L\'{e}vy random matrices, and demonstrate a novel feature 
of {\it universal scaling in the tails}. Our conclusions are given in section 5.

\section{Resolvents of Free L\'{e}vy Variables}

The fundamental problem in random matrix theory is to find the
distribution of eigenvalues $\lambda_i $ in the large $N$ (size of
$M$) limit, i.e
\be
\rho(\lambda)= \frac{1}{N} \left\la \sum _{i=1}^N 
	\delta (\lambda - \lambda_i)\right\ra =\frac 1N \,
\left\la\,{\rm Tr}\,\delta(\lambda -M)\right\ra
\label{spect-h}
\ee
where the averaging is carried using some pertinent measure
\be
e^{-N \,{\rm Tr }\,V(M)}  \,\,dM\,\,.
\label{MEASURE}
\ee
The potential $V(M)$ does not have to be a polynomial or to
have even a power series expansion, as usually assumed. Indeed, 
we will show below that for L\'{e}vy distributions the
potential is not analytic in $M$. In general, 
it is convenient to introduce the Green's function
\be
G(z)=\frac{1}{N} \left\la {\rm Tr}\, \frac{1}{z-M}\right\ra\,\,.
\label{green}
\ee
The eigenvalue distribution follows from the discontinuity
of $G(z)$ along the real axis, i.e. 
$\rho (\lambda ) =-\,{\rm Im}\,G(\lambda +i0)/\pi$.
In this paper we are interested in stable random matrix ensembles
i.e. ensembles in which the eigenvalue distribution of the sum
$M=M_1+M_2$, 
is the same (up to a shift and/or rescaling) as for each individual 
matrix $M_1$ and $M_2$. If the second moment exists, then the ensemble
is Gaussian. In the opposite case there exist, however, alternate
ensembles which are the matrix analogues of the classical
one-dimensional L\'{e}vy distributions. 

The notion of addition laws of the type $M=M_1+M_2$ has been
analyzed in the context of the theory of free random variables (FRV), a
generalization by Voiculescu~\cite{VOIC} of classical probability theory to a
noncommutative setting. A novel twist in the theory of FRV
came from the realization, that 
the abstract concepts of operator algebras and free random variables
can have an explicit realization in terms of large random matrices. 
Hence  FRV techniques provide a novel and powerful way of analyzing
the spectra of random matrices~\cite{ZEE,BLUES}. 
In this framework the determination of stable {\em eigenvalue 
distributions} can be performed algebraically in a very general
setting. This has been done by Bercovici and
Voiculescu~\cite{BV}. However the drawback of such an abstract
approach is that the explicit random matrix ensembles corresponding to these
distributions are unknown. The aim of this paper is to fill this gap,
as the knowledge of explicit stable random matrix ensembles with power
law tails might be interesting for the various applications we have cited.

The remarkable achievement by Bercovici and Voi\-cu\-les\-cu~\cite{BV}
is an explicit derivation of all R-transforms (defined by the equation
$R(G(z))=z-1/G(z)$ where $G(z)=\langle 1/(z-\lm) \rangle$) for all free 
stable distributions, without recourse to a matrix realization.
Indeed, Bercovici and Voiculescu have found that
$R(z)$ can have the trivial form $R(z)=a$ 
or~:
\be
R(z) = a + b z^{\alpha-1}
\ee
where $0<\alpha<2$, $a$ is a real shift parameter,
and $b$ is a parameter which can be related to the
slope $\alpha$, skewness $\beta$, and range $\gamma$ of
the standard parameterization of stable distribution \cite{BV,PATA}~:
\indent{\be
  b = \left\{ \begin{array}{cl}
	\gamma\ e^{i(\frac{\alpha}2-\!1)(1\!+\!\beta)\pi} 
		{\rm ~~for~~} 1 <\alpha<2 \nonumber \\
	\gamma\ e^{i[\pi+\frac{\alpha}2(1\!+\!\beta)\pi]} 
		{\rm ~~for~~} 0 <\alpha <1 \nonumber
	\end{array} \right. \, .
\ee}
In the marginal case: $\alpha=1$, $R(z)$ reads~:  
\be
R(z) = a - i \gamma(1+\beta) -
\frac{2\beta\gamma}{\pi} \ln \gamma z
\ee
The branch cut structure of $R(z)$ is chosen in such a way that the
upper complex half plane is mapped to itself.
Recalling that $R=z-1/G$ in the large $N$ limit, 
one finds that for the trivial case $R(z)=a$,
the resolvent~: $G(z)=(z-a)^{-1}$ 
and the spectral distribution 
$\rho(\lambda)=\delta(\lambda-a)$.
Otherwise, on the upper half-plane, the resolvent fulfills 
an algebraic equation
\be
bG^{\alpha}(z)-(z-a)G(z)+1=0\,\,,
\label{Levygreen}
\ee
or in the marginal case ($\alpha=1$)~:
\be
\bigg(\!z\!-a+i\gamma(1\!+\!\beta)\!\bigg) G(z) +\frac{2\beta\gamma}{\pi}
G(z)\ln \gamma G(z) - 1\! = \!0.
\ee
On the lower half-plane $G(\bar{z})=\bar{G}(z)$ \cite{BV}.
The solution of the latter equation
will not be discussed here, except for 
$\beta=0$ for which it simplifies to
\be
G(z)=\frac{1}{z-(a-i\gamma)}\,\,.
\ee
Thus in this case the spectral density has the form 
of a Cauchy distribution~: (L\'{e}vy with $\alpha=1$)~:
\be
\rho(\lambda)=\frac{1}{\pi}\frac{\gamma}{(\lambda-a)^2 +\gamma^2}\,\,.
\ee

For $\alpha=1/4,1/3,1/2, 2/3,3/4,4/3,3/2$ and $2$ 
the algebraic equation (\ref{Levygreen}) are exactly solvable.
The corresponding spectral functions are given 
by appropriate solutions of the quadratic, cubic ( Cardano)
or quartic (Ferrari) equations. The case $\alpha=2$ corresponds 
to the known Gaussian case.  
We note that the spectral density for slope parameter $1/\alpha$ 
follows from
\be
   \rho_{\alpha,\beta}(x) = x^{-1-\alpha} 
	\rho_{\frac1{\alpha},\beta'}(x^{-\alpha})
\ee
for $1<\alpha<2$ and $\beta'=(\alpha-1)-(2-\alpha)\beta$~\cite{PATA}, 
again by analogy to the well known duality relations for one-dimensional
stable distributions~\cite{ZOLO}.
%

Despite the fact that one-dimensional  L\'{e}vy distributions are unknown
explicitly (the exception being the three cases $\alpha=1/2,1,2$), the
corresponding resolvents for free random L\'{e}vy matrices
obey simple multivalued equations. 
Eq.~(\ref{Levygreen}) has an asymptotic solution for 
$\beta=0$, $\gamma=1$ and
$\lambda\to 0$,
\be
  \rho(\lambda) = \frac1{\pi} \left( 1-\frac{3-\alpha}{2\alpha^2}
\lambda^2 + \cdots\right) \,
\ee
and one as $\lambda\to\infty$,
\be
  \rho(\lambda) = \frac{\sin{\alpha \frac{\pi}2}}{\pi}\
	\lambda^{-\alpha-1} \,.
\ee
The scaling connecting the distribution with unit scale to the one
with scale $\gamma$,
\be
  \rho_{\gamma}(\lambda) = \gamma^{-\frac1{\alpha}}\ 
	\rho( \gamma^{\frac1{\alpha}}\lambda) \,.
\ee

\section{Random Matrix Stable L\'{e}vy Ensembles}

In our quest for a random matrix realization of the above free stable
distributions we will look for hermitian~\footnote{The other two
possibilities, symmetric and skew-symmetric ones, could be 
considered along the lines presented in this work.} random matrix
ensembles with the measure (\ref{MEASURE}),
such that the corresponding $\rho (\lambda) $ coincides with the large
$N$ limit of the
mean eigenvalue distribution. From the theory of FRV
such ensembles will be automatically stable with respect to addition.
Here the potential $V(M)$ is not restricted to be a polynomial (or a
power series) but can involve logarithms etc.
Note that in particular we require unitary (orthogonal) symmetry $M\to
UMU^\dagger$ of the ensemble. 

The standard procedure of diagonalizing $M\to U\Lambda U^\dagger$ and
integrating out $U$ gives rise to the standard 
joint probability distribution for the eigenvalues
\eq
\label{e.cg}
\rho(\lm_1,.\!.\!.,\lm_N) \prod_i d\lm_i = \prod_i d\lm_i
e^{-NV(\lm_i)} \prod_{i<j} (\lm_i-\lm_j)^2 \,\,.
\eqx  
For large $N$, the corresponding partition function can be analyzed in
terms of the Coulomb gas action with a continuous eigenvalue
distribution $\rho(x)$ as originally suggested by
Dyson~\cite{DYSON}. Specifically, 
\eq
\frac{S(\rho)}{N^2}=\int \! d\lambda \rho(\lambda) V(\lambda) - \int \!
d\lambda d\lambda^{'} \ln |\lambda\!-\!\lambda^{'}|
\rho(\lambda) \rho(\lambda^{'}) \, .
\eqx
Two functional differentiations yield,
\be
V'(\lambda)=2\,  {\rm PV} \int d\lambda^{'} \frac{\rho(\lambda^{'})}{\lambda
-\lambda^{'}}\,\,,
\ee
where PV denotes the principal value of the integral.  The
knowledge of 
 $V(\lambda)$ plus the boundary conditions on the singular integral 
equation allows to find the spectral function~\cite{MUSHKALISHVILI}.
Conversely, from the knowledge of the spectral function we can deduce 
the shape of the potential, hence the weight of the probability 
distribution for the matrix ensemble. This is the route for implementing
the FRV results on random matrix ensembles as we now show.

Using the formula $1/(\lambda\pm i\eps)=PV\, \f{1}{\lambda}\mp 
i\pi \dl(\lambda)$ 
we get a formula for $V'(\lambda)$ in terms of the real part of the
resolvent on the cut $V'(\lambda)=2\, \mbox{\rm Re }\, G(\lambda)$.
We can now use the resolvents found in section 2 to reconstruct the
potential and hence to explicitly define a random matrix realization of
the free stable L\'{e}vy ensembles. Below, we reconstruct the  pertinent 
measures for the Cauchy ($\alpha=1, \beta=0$) and Smirnov 
($\alpha=1/2, \beta =\pm 1$) distributions. In the cases  $\alpha\neq 1/4,
1/3, 1/2, 2/3,3/4,4/3,3/2$ such construction can be performed numerically.
General results can be achieved without these specifics as we will show
below.

In the case of the Cauchy distribution, elementary integration gives
$V'(\lambda)=\gamma 2\lambda/(\lambda^2+b^2)$, hence the potential
$V(\lambda) =\gamma \ln (\lambda^2+b^2)$.
In the case of the Smirnov distribution, the Green's function follows from
\eq
\f{-i}{\sqrt{G}}+\f{1}{G}=z
\eqx
which can be easily solved to yield
\eq
G(z)=\frac{2z-1 -i \sqrt{4z-1}}{2z^2}\,\,.
\eqx
Evaluating the real part gives $V'(\lambda)=2/\lambda -1/\lambda^2$,
so that
\eq
\label{e.lsvm}
V(M)= \exp\left(-N\,{\rm Tr} \left(  \f{1}{M} + 2\, {\rm ln} \, M\right)
\right) \,\,.
\eqx
The spectral function follows 
from (3) and (21)  and 
reads $\rho(\lambda)=1/(2\pi)\sqrt{4\lambda-1}/\lambda^2$.

In general, the asymptotic form of the potential
for large eigenvalues reads
\eq
\label{e.vasympt}
V( \lambda)=2\,{\rm ln}\,\lambda-2\,\f{1}{\alpha}\, {\rm Re}\, b
\,\frac{1}{\lambda^{\alpha}}  + ...
\eqx
In all L\'{e}vy cases  the $2\,{\rm ln}\,\lambda$ contribution
in the potential is fixed, and generate in the measure a determinant
with also fixed power $-2N$. A deviation from 2 can be shown to lead to a finite 
support of eigenvalues. The coefficient of the second term in the
potential can vanish in some
notable cases like e.g. for the Smirnov ensemble. 
In the next section we will analyze in greater detail the stable
L\'{e}vy random matrix ensembles defined above.

\section{Orthogonal polynomials}

Both the Cauchy and Smirnov ensembles can be studied analytically at
finite $N$ using the orthogonal polynomial method. It turns out that
the above ensembles have some unexpected new features which do not
appear in the classical case when $V(M)$ is a polynomial in $M$.
The orthogonal polynomials associated with a random matrix ensemble
defined by $V(\lm)$ satisfy the orthogonality relation
\eq
\label{e.orthgen}
\int d\lm e^{-NV(\lm)} P_n(\lm)P_m(\lm)=\dl_{nm}
\eqx
Then the exact expression for arbitrary $N$ for the density of
eigenvalues is
\eq
\label{e.rhopol}
\rho(\lm)=e^{-NV(\lm)} \f{1}{N} \sum_{i=0}^{N-1} P_n^2(\lm)
\eqx
All higher correlation functions can be recovered from the kernel
\eq
k(\lm_1,\lm_2)=e^{-\f{N}{2}(V(\lm_1) +V(\lm_2))} \f{1}{N}\sum_{i=0}^{N-1}
P_n(\lm_1) P_n(\lm_2)
\eqx


For the Cauchy random matrix ensemble the orthogonality relation
(\ref{e.orthgen}) reads
\eq
\int d\lm (\lm^2+1)^{-2N} P_n(\lm)P_m(\lm)=\dl_{nm}
\eqx
We see that in contrast to the classical case only a {\em finite}
number of orthogonal polynomials exist. These are explicitly given by
Jacobi polynomials analytically continued to complex
parameters~\footnote{ After completing the paper, we noticed 
that a similar construction was recently used in~\cite{WITTE}.}.
 Indeed
\eq
P_n(x)\!=\!\left(\f{(1\!+\!n\!-2\!N)_n}{2^{2n} n!} \sqrt{\pi}
\f{\Gamma\left(N\!-\!n\!-\!\f{1}{2}\right)}{\Gamma(N\!-\!n)} 
\right)^{\!\!-\f{1}{2}} \!\! i^n J^{-\!N\!,-\!N}_n(ix)
\eqx
A second surprise comes from the fact that the eigenvalue distribution
(\ref{e.rhopol}) is exactly equal to
\eq
\rho(\lm)=\f{1}{\pi}\f{1}{\lm^2+1}
\eqx
and does {\em not} depend on $N$. There are no finite $N$ corrections
whatsoever to the spectral distribution. In particular the classical
short distance oscillations in the spectral density characteristic for 
RMM are absent.
The Cauchy ensemble has, however, nontrivial $N$-dependent 2-point
correlation functions.

 
The L\'{e}vy-Smirnov ensemble can be analyzed starting from the CG
distribution (\ref{e.cg}) with the measure (\ref{e.lsvm})
\eq
\prod_i d\lm_i \left(\f{e^{-N/\lm_i}}{\lm^{2N}}\right)
\prod_{i<j} (\lm_i-\lm_j)^2
\eqx
A change of variables $\lm_i=1/x_i$ leads to 
\eq
\prod_i dx_i \left(e^{-Nx_i}\right)
\prod_{i<j} (x_i-x_j)^2
\eqx
This is readily analyzed in terms of Laguerre polynomials
similar to Chiral Unitary Ensembles (chGUE) \cite{ZV}. Indeed the appropriate
polynomials are
\eq
P_n(x)=\sqrt{N} L^0_n(Nx)\,\,.
\eqx
The eigenvalue density (\ref{e.rhopol}) can be rewritten, using the
Christoffel-Darboux identity, as
\eq
\rho(x)\!=\!N \! e^{-N\!x} 
\!\left( L^0_{N\!-\!1}(N\!x)L_{N\!-\!1}^1(N\!x) - 
L^0_{N}(N\!x)L_{N\!-\!2}^1(N\!x) 
\right)
\eqx
In particular we note that there exist a well defined microscopic
limit which corresponds to expressing the eigenvalue density in terms
of $x=s/N^2$ (i.e. on the scale of the eigenvalue spacing). The relevant
eigenvalue density $(1/N)\rho(s/N^2)$ is then given by
\eq
\rho(s)=J^2_0(2\sqrt{s})+J^2_1(2\sqrt{s})
\eqx
Going back to the original variables the microscopic region
corresponds to the region of {\em large} eigenvalues $\lm=N^2\Lm$ in
the power-like
tail. For these {\em large} eigenvalues we therefore observe chGUE-like
oscillations
\eq
\label{e.macroosc}
\rho(\Lm)=\f{1}{\Lm^2}\left\{ J_0^2\left(\f{2}{\sqrt{\Lm}}\right)+ 
J_1^2\left(\f{2}{\sqrt{\Lm}}\right) \right\}
\eqx
Moreover we expect these oscillations to be {\em universal} in the following
sense. A generic modification of the LS potential of the form
\eq
V(\lm)=2\log \lm +\f{1}{\lm}+\f{g_2}{\lm^2}+\f{g_3}{\lm^3}+\ldots
\eqx 
will not change the oscillation pattern (\ref{e.macroosc}). This
follows from the results in~\cite{DAMGAARD}
after the change of variables $\lm \to x=1/\lm$. The coefficient in
front of the logarithm cannot be changed, for otherwise the eigenvalue
support becomes finite and the power like tails disappear altogether.
The coefficients of the $1/\lm$ term (and higher $g_i$'s) only affect
the length scale of the  universal oscillations.


In the general case when the potential is of the form
(\ref{e.vasympt}), and the asymptotic behavior of $\rho(\lm)$ is like
$1/\lm^{1+\al}$, the mapping $\lm \to x=1/\lm$ gives an effective
potential 
\eq
V(x)\sim -2\f{1}{\alpha}\, {\rm Re}\, b \,\,x^\al
\eqx
and an eigenvalue distribution $\rho(x)\sim x^{\al-1}$. General arguments
\cite{CRITICAL} show that the resulting eigenvalue spacing $1/N^{1/\alpha}$
yields a microscopic distribution in the limit of $N\to \infty$ 
with $s=x N^{1/\alpha}$ fixed. 
The pertinent orthogonal polynomials should satisfy
\eq
\int dx e^{-N|x|^\alpha} P_n(x)P_m(x)=\dl_{nm} \, .
\eqx 
All this is very reminiscent of multicritical microscopic scaling and
universality \cite{CRITICAL,DAMGAARD2,HB} 
in the classical random matrix case but
here the ``multicritical classes'' are labeled by a {\em real}
parameter $\alpha$ and not by an {\em integer}. A thorough
investigation of this new regime seems to be very interesting.

\section{Summary}

We have explicitly constructed matrix realizations of free
random variables, with potential applications to a number of
stochastic phenomena. This opens  several new venues for 
applying FRV calculus to L\'{e}vy processes, including convolution, 
multiplication and addition of deterministic matrix-like 
entries and other generalizations. Using the Coulomb gas analogy, 
we have shown that the exact matrix measure in the case of power-like
spectra is non-local (involves determinants).
The construction exhibits several non-trivial and novel 
features, among which the most interesting ones are a 
universal behavior in the tails of the distributions and an 
unusual large $N$ scaling.
The expected microscopic eigenvalue distribution defines a new universal regime 
and  represents a generalization
of the multicritical  scaling discussed in \cite{CRITICAL,DAMGAARD2,HB}. 
Several of these issues, as well as practical applications 
of our results will be discussed in subsequent work.

\vskip 2cm
{\bf Acknowledgments:}
\vskip .5cm
This work was supported in part by the US DOE grants DE-FG02-88ER40388
and DE-AC02-98CH10886,
the Polish Government Projects (KBN) 2P03B 00814, 
2P03B 01917, the Hungarian FKFP grant 220/2000, and 
the EC IHP network ``Discrete Random Geometry'', 
grant HPRN-CT-1999-00161.

\end{document}